\documentclass[pdflatex,sn-mathphys]{sn-jnl}

\RequirePackage{color}
\definecolor{MyDarkGreen}{rgb}{0.02,0.60,0.06}

\usepackage{ulem}

\jyear{2021}%

\theoremstyle{thmstyleone}%
\theoremstyle{thmstyletwo}%

\theoremstyle{thmstylethree}%

\newcommand{\de}[2]{{\frac{\partial #1}{\partial #2}}}

\begin{document}


\title{Influence of anisotropic matter on the Alcubierre metric and other related metrics: revisiting the problem of negative energy}


\author[1,2]{\fnm{Gabriel} \sur{Abell\'an}}\email{gabriel.abellan@ciens.ucv.ve}
\email{gabriel@astrumdrive.com}

\author*[1,2]{\fnm{Nelson} \sur{Bolivar}}\email{nelsonbolivar@cnea.gob.ar}
\email{nelson@astrumdrive.com}

\author[1,4]{\fnm{Ivaylo} \sur{Vasilev}}\email{ivaylo@astrumdrive.com}

\affil[1]{\orgdiv{Astrum Drive Technologies}, \orgaddress{\city{Dallas Pkwy Unit 120 B, Frisco, TX.}, \postcode{ 75034},  \country{USA}}}

\affil[2]{\orgdiv{Departamento de F\'isica}, \orgname{Facultad de Ciencias,
Universidad Central de Venezuela}, \orgaddress{\street{Av. Los Ilustres}, \city{Caracas}, \postcode{1041-A},  \country{Venezuela}}}

\affil[4]{ \orgname{Technical University of Sofia}, \orgaddress{ \city{Sofia}, \postcode{1000}, \country{Bulgaria}}}


\abstract{
Negative energy scenarios are the most widely studied for the warp metric. In fact, the prevailing view in the community so far has been that the warp metric necessarily has negative energies. In this work it is shown that the issue of negative energy densities associated with the Alcubierre warp metric with a general form function and similar metrics can be addressed when the whole non--vacuum Einstein equations of the system are examined. To this end, we have considered matter content in the form of anisotropic fluids. 
We have succeeded in writing the Einstein equations in such a way that some general constraints on the material content become evident. This means that, in rectangular coordinates, the energy density depends necessarily on the tangential pressures of the fluid. 
For matter such as dust or isotropic fluids we find that density and other related quantities become identically zero. This makes the negative energy problem spurious. 
It is also revealed that constructing Alcubierre-based metrics using cylindrical and spherical coordinates results in a system of equations that are amenable to more systematic analysis. The field equations constrain the dependence of the form function and how this impacts the matter content. 
In all cases we determine that energy density is not mandatory negative, despite the recurrent claims in the literature.
This result prompts a reevaluation of the negative energy requirements and underscore the importance of cylindrical and spherical type-warps to demonstrate that negative energy density is not an intrinsic unavoidable feature of warp drives.
}




\maketitle

\section{Introduction}\label{sec-01}
In 1994 the warp drive metric was proposed as a way of modelling displacement at superluminal velocities \cite{Alcubierre:1994tu}. This raised immediate interest and continues to be of interest to the community. 
One of the aspects that has been widely discussed is that in studying the expression for the energy density needed to support the warp metric, it was found that it must be negative. It is only recently, however, that a thorough examination of the Einstein equations as a whole has been carried out for this type of metric. 

All these aspects, as well as the implementation of energy conditions, have been extensively studied over the years \cite{Visser1995lwet,Everett:1995nn,Krasnikov:1995ad,Hiscock:1997ya,Everett:1997hb,Visser:1997tq,Olum:1998mu,Low:1998uy,Pfenning:1998ua,Gonzalez-Diaz:1999vtr,Visser:1999de,Clark:1999zn,VanDenBroeck:1999sn,VanDenBroeck:1999xs,Barcelo:2000zf,Barcelo:2002bv,Lobo:2002zf,Lobo:2004an,Lobo:2004wq,Curiel:2014zba,Lobo2017,Alcubierre:2017kqf,Kontou:2020bta,Bobrick:2021wog,Lentz:2021rzh,Santiago:2021aup,Schuster:2022ati,Guo:2023kow}.

As is well known, to study problems in general relativity we must consider the Einstein field equations
\begin{equation}\label{e-eq}
	G_{\mu\nu} = 8\pi T_{\mu\nu}\;,
\end{equation}
which relate the geometry of spacetime to the material content. Typically, the problem of energy densities related to the warp drive has been studied by looking at the geometry using the relation
\begin{equation}
	G_{\mu\nu}u^{\mu}u^\nu = 8\pi T_{\mu\nu} u^{\mu}u^\nu
	\hspace{.5cm} \longrightarrow \hspace{.5cm}
	\rho = \frac{1}{8\pi} G_{\mu\nu}u^\mu u^\nu \;,
\end{equation}
with $u^\mu$ characterising a free-falling observer. However, as mentioned above, rarely has there been a more exhaustive exploration of the full matter content and even more of Einstein's equations in their integrity.
Recently, Santos et al. carried out a study of the Einstein equations and searched to discern the matter content that could be consistent with the warp metric \cite{Santos-Pereira:2020puq,Santos-Pereira:2021mqp,Santos-Pereira:2021mrr,Santos-Pereira:2021rsr,Santos-Pereira:2021sdm,Abellan2023}. To do so, they heuristically suggested a form for the energy--momentum tensor and made the analysis. Unfortunately, the structure of the Einstein equations is so complex that the analytical work of drawing conclusions is always extremely difficult.
This is why in this work we intend to perform a revision of the original Alcubierre warp with general form function in Cartesian coordinates and other related warp metrics with other symmetries, studying in detail which are the restrictions imposed by the Einstein equations as well as the limitations related to the material content. In this sense we will consider as a source an anisotropic fluid which contains sufficient flexibility to accommodate the cases of isotropic fluid and dust. All these cases are of interest for the characterization of possible sources of matter.

We propose to revisit the Alcubierre metric with general form function by examining in more detail the structure of the Einstein equations and the matter content that supports the warp spacetime. We further propose to examine how the constraints in the equations determine conditions for the energy density and how these conditions change as we modify the symmetry of the residual flat space proper to any warp metric.

In order to analyze the generality of these results, in section \ref{sec-02} we will study the original alcubierre metric with a general form function, looking in depth at the material content able to sustain this configuration. 
In order to do this we propose a rigorous writing for the energy momentum tensor describing anisotropic matter. We will then see that by writing the Einstein equations in a convenient form, some restrictions become evident which are of central importance for drawing conclusions about the energy density.
 
Then in section \ref{sec-03}, we consider an Alcubierre-type warp where the residual flat geometry appearing in the line element is written in cylindrical coordinates. 
In this section will focus on examining the warp in the $z$-direction and identifying the specific matter requirements for sustaining this metric. 
We further explore some formal similarities with the Cartesian case above.
Subsequently, in sections \ref{sec-04} and \ref{sec-05} we will perform analogous procedure to the previous one but warping in the cylindrical radial coordinate and in the spherical radial coordinate respectively. In all sections our main interest is to determine how the various constraints imposed by the entire set of Einstein equations and the matter content determine the energy density. Finally we will give some conclusions in section \ref{sec-06}.

\section{Alcubierre warp drive}\label{sec-02}
In this section, we review the general elements of Alcubierre's original article \cite{Alcubierre:1994tu}. The proposed line element was
\begin{equation}\label{al-1}
	ds^2 = -dt^2 + (dx - \beta dt)^2 + dy^2 + dz^2\;.
\end{equation}
This line element was inspired by 3+1 formalism. The metric in matrix form is given by
\begin{equation}\label{metric-alc}
	g_{\mu\nu} =
	\left[ \begin{array}{cccc}
		-(1-\beta^2) & -\beta & 0 & 0 \\
		-\beta & 1 & 0 & 0 \\
		0 & 0 & 1 & 0 \\
		0 & 0 & 0 & 1
	\end{array} \right] .
\end{equation}
In Alcubierre's original article the quantities involved are defined as follows
\begin{equation}\label{al-2}
	\beta = v_s(t) f(r_s) \;, \hspace{.6cm} 
	r_s = \sqrt{(x-x_s(t))^2 + y^2 + z^2} \;, \hspace{.6cm}
	v_s(t) = \frac{dx_s(t)}{dt}\;,
\end{equation}
where the function $f(r_s)$ is given by
\begin{equation}\label{al-3}
	f(r_s) = \frac{\tanh{[\sigma(r_s + R)]} - \tanh{[\sigma(r_s - R)]}}{2\tanh{(\sigma R)}}\;,
\end{equation}
with $\sigma$ and $R$ positive parameters. From equation (\ref{al-2}) it is clear that $\beta$, as proposed by Alcubierre, should be a function of all coordinates $\beta(t,x,y,z)$.

In order to study the condition on energy density, Alcubierre considered an Eulerian (free falling) observer characterized by
\begin{equation}\label{al-4}
	u_\mu = (-1,0,0,0)\;, \hspace{.7cm}
	u^\mu = (1,\beta,0,0)\;.
\end{equation}
Note that this is a timelike vector $u^\mu u_\mu = -1$. Using these equations and the well-known relation $T^{\mu\nu}u_\mu u_\nu = \rho$ he found an expression for the energy density
\begin{equation}\label{al-5}
	T^{00} = \frac{G^{00}}{8\pi} = 
	-\frac{1}{32\pi} v_s^2 \Bigg(\frac{df}{dr_s}\Bigg)^2 \frac{y^2 + z^2}{r^2_s} = \rho\;.
\end{equation}
For further analysis, it is convenient to express the above equation in terms of the $\beta$ function
\begin{equation}\label{al-6}
	\rho = 
	-\frac{1}{32\pi} \Bigg[ \left( \frac{\partial\beta}{\partial y} \right)^2 
	+ \left( \frac{\partial\beta}{\partial z} \right)^2\Bigg]\;.
\end{equation}
Regardless of whether it is expressed as (\ref{al-5}) or (\ref{al-6}), it is possible to observe the same behavior that Alcubierre noted and which has been the subject of extensive discussion up to now, namely, that the warp metric applies only to matter with negative energy density

\subsection{Einstein's equations and constraints}

Alcubierre primarily addressed the geometrical aspects when introducing the warp metric. However, a comprehensive examination of the issue requires considering the matter content as well. To achieve this, we suggest using an anisotropic fluid as a source
\begin{equation}\label{temunuaniA}
	T_{\mu \nu} = (\rho + p_x) u_\mu u_\nu + p_x g_{\mu \nu} + (p_y - p_x) s_\mu s_\nu + (p_z - p_x) t_\mu t_\nu \;.
\end{equation}

The 4--vector $u_\mu$ is defined in (\ref{al-4}) while $s_\mu$ and $t_\mu$ are spacelike 4-vectors pointing in the $y$-direction and $z$-direction respectively. They are given by
\begin{equation}\label{4vectors-al}
	s_\mu = (0,0,1,0) \;, \hspace{.8cm}
	t_\mu = (0,0,0,1) \;.
\end{equation}

These vectors satisfy the relations $s^\mu s_\mu = 1$, $t^\mu t_\mu = 1$ and $u^\mu s_\mu = u^\mu t_\mu = 0$.

In matrix form the energy--momentum tensor is written as
\begin{equation} \label{tmunumatrixal}
	T_{\mu \nu} =    \left[
	\begin{array}{cccc}
		\rho + \beta^2 p_x  & -\beta p_x & 0 & 0 \\
		-\beta p_x & p_x & 0 & 0 \\
		0 & 0 & p_y & 0 \\
		0 & 0 & 0 & p_z \\
	\end{array}
	\right]
\end{equation}
Here $\rho$ corresponds to the energy density, $p_x$ is the normal pressure and $p_y$ and $p_z$ are the tangential pressures. Note that for isotropic matter we should put $p_x = p_y = p_z = p$ while for dust we set $p_x = p_y = p_z = 0$.

Using the metric (\ref{metric-alc}) and the energy--momentum tensor (\ref{tmunumatrixal}) we can write Einstein's equations (\ref{e-eq}) explicitly
\begin{eqnarray}
	-\frac{1}{4}(1+3\beta^2) \Bigg[ \left( \frac{\partial\beta}{\partial y} \right)^2 
	+ \left( \frac{\partial\beta}{\partial z} \right)^2\Bigg]
	- \beta \left( \frac{\partial^2\beta}{\partial y^2} 
	+ \frac{\partial^2\beta}{\partial z^2} \right) 
	= 8\pi (\rho + \beta^2 p_x)\;,  & \label{e-1} \\
	\frac{3}{4}\beta \Bigg[ \left( \frac{\partial\beta}{\partial y} \right)^2 
	+ \left( \frac{\partial\beta}{\partial z} \right)^2\Bigg]
	+ \frac{1}{2} \left( \frac{\partial^2\beta}{\partial y^2} 
	+ \frac{\partial^2\beta}{\partial z^2} \right)
	= -8\pi\beta p_x \;, \;\;\;\;\;\;\;\; & \label{e-2} \\
	-\frac{3}{4} \Bigg[ \left( \frac{\partial\beta}{\partial y} \right)^2 
	+ \left( \frac{\partial\beta}{\partial z} \right)^2\Bigg]
	= 8\pi p_x \;, \;\;\;\;\;\;\;\;\;\;\;\;\;& \label{e-3} \\
	-\frac{1}{2}\frac{\partial^2\beta}{\partial x \partial y}
	-\frac{1}{2}\beta \left( 
	2 \frac{\partial\beta}{\partial x} \frac{\partial\beta}{\partial y}
	+ \beta \frac{\partial^2\beta}{\partial x \partial y} 
	+ \frac{\partial^2\beta}{\partial t \partial y}
	\right) 
	= 0 \;, \;\;\;\;\;\;\;\;\;\;\;\;\;\;\;\;\;\;\; & \label{e-4} \\
	\frac{1}{2} \left( 
	2 \frac{\partial\beta}{\partial x} \frac{\partial\beta}{\partial y}
	+ \beta \frac{\partial^2\beta}{\partial x \partial y} 
	+ \frac{\partial^2\beta}{\partial t \partial y}
	\right) 
	= 0 \;, \;\;\;\;\;\;\;\;\;\;\;\;\;\;\;\;\;\;\; & \label{e-5} \\
	- \left[
	\left(\frac{\partial\beta}{\partial x}\right)^2
	+ \beta \frac{\partial^2\beta}{\partial x^2} 
	+ \frac{\partial^2\beta}{\partial t \partial x}
	\right]
	+ \frac{1}{4} \Bigg[ \left( \frac{\partial\beta}{\partial y} \right)^2 
	- \left( \frac{\partial\beta}{\partial z} \right)^2\Bigg]
	= 8\pi p_y \;, \;\;\;\;\;\;\;\;\;\;\;\;\;&  \label{e-6} \\
	-\frac{1}{2}\frac{\partial^2\beta}{\partial x \partial z}
	-\frac{1}{2}\beta \left( 
	2 \frac{\partial\beta}{\partial x} \frac{\partial\beta}{\partial z}
	+ \beta \frac{\partial^2\beta}{\partial x \partial z} 
	+ \frac{\partial^2\beta}{\partial t \partial z}
	\right) 
	= 0 \;, \;\;\;\;\;\;\;\;\;\;\;\;\;\;\;\;\;\;\; & \label{e-7} \\
	\frac{1}{2} \left( 
	2 \frac{\partial\beta}{\partial x} \frac{\partial\beta}{\partial z}
	+ \beta \frac{\partial^2\beta}{\partial x \partial z} 
	+ \frac{\partial^2\beta}{\partial t \partial z}
	\right) 
	= 0 \;, \;\;\;\;\;\;\;\;\;\;\;\;\;\;\;\;\;\;\; & \label{e-8} \\
	\frac{1}{2} \frac{\partial\beta}{\partial y} \frac{\partial\beta}{\partial z}
	= 0 \;, \;\;\;\;\;\;\;\;\;\;\;\;\;\;\;\;\;\;\; & \label{e-9} \\
	- \left[
	\left(\frac{\partial\beta}{\partial x}\right)^2
	+ \beta \frac{\partial^2\beta}{\partial x^2} 
	+ \frac{\partial^2\beta}{\partial t \partial x}
	\right]
	- \frac{1}{4} \Bigg[ \left( \frac{\partial\beta}{\partial y} \right)^2 
	- \left( \frac{\partial\beta}{\partial z} \right)^2\Bigg]
	= 8\pi p_z \;. \;\;\;\;\;\;\;\;\;\;\;\;\; & \label{e-10} 
\end{eqnarray}

The equations (\ref{e-1})--(\ref{e-10}) constitute a formidable coupled system of non--linear partial differential equations. It is not the purpose of this article to solve these equations but rather to point out that there are some restrictions on the $\beta$ function that can be derived directly from them.

From the previous relations, we extract the following constraints
\begin{eqnarray}
	\mbox{from (\ref{e-5}) into (\ref{e-4})\,:} 
	\hspace{.7cm} \frac{\partial^2\beta}{\partial x \partial y} 
	= 0 \;,\;\;\;\;\;\;\;\;\;\;\;\;\;\;\;\;\;\;\;\;\;\;\;\;\;\;\;\;\;\;\;\;\;\;\;\;\;\;\;\;\;\;\;\;\;\;\;\;\;\;\; \label{cons-1} \\
	\mbox{from (\ref{e-8}) into (\ref{e-7})\,:} 
	\hspace{.7cm} \frac{\partial^2\beta}{\partial x \partial z} 
	= 0\;, \;\;\;\;\;\;\;\;\;\;\;\;\;\;\;\;\;\;\;\;\;\;\;\;\;\;\;\;\;\;\;\;\;\;\;\;\;\;\;\;\;\;\;\;\;\;\;\;\;\;\; \label{cons-2} \\
	\mbox{from (\ref{e-9})\,:} 
	\hspace{.7cm} \frac{\partial\beta}{\partial y} = 0 \hspace{.3cm}
	\mbox{or} \hspace{.3cm} 
	\frac{\partial\beta}{\partial z} 
	= 0\;, \;\;\;\;\;\;\;\;\;\;\;\;\;\;\;\;\;\;\;\;\;\;\;\;\;\;\;\;\;\;\;\;\;\;\;\;\;\;\;\;\;\;\;\;\;\;\;\;\; \label{cons-3} \\
	\mbox{from (\ref{e-6}) $+$ (\ref{e-10})\,:} 
	\hspace{.7cm} \left( \frac{\partial\beta}{\partial x} \right)^2
	+ \beta \frac{\partial^2\beta}{\partial x^2}
	+ \frac{\partial^2\beta}{\partial t \partial x}
	= -4\pi (p_y+p_z) \;, \;\;\;\;\;\;\label{cons-4} \\
	\mbox{from (\ref{e-6}) $-$ (\ref{e-10})\,:} 
	\hspace{.7cm} \frac{1}{2} \left[ \left( \frac{\partial\beta}{\partial y} \right)^2 
	- \left( \frac{\partial\beta}{\partial z} \right)^2 \right]= 8 \pi (p_y-p_z) \;. \;\;\;\;\;\;\;\;\;\;\;\; \label{cons-5}
\end{eqnarray}
Of all the constraints found, the one given by (\ref{cons-3}) is critical. 

No matter the choice in conditions (\ref{cons-3}), one of the constrains (\ref{cons-1}) or (\ref{cons-2}) is immediately satisfied. In turn, it impacts on equation (\ref{cons-5}), which is further simplified.

As a consequence of equation (\ref{cons-3}), the $\beta$ function is restricted to depend only on $(t,x,y)$ or $(t,x,z)$ . This leads to a simpler system, which is reflected in the energy-momentum tensor representing the matter content.
In particular, it can be seen that this necessarily implies that either we have
\begin{equation}\label{nullrhoplus}
	\rho = 
	-\frac{1}{32\pi} \Bigg[ \left( \frac{\partial\beta}{\partial y} \right)^2 
	+ \left( \frac{\partial\beta}{\partial z} \right)^2\Bigg]         
	\hspace{.7cm} \longrightarrow \hspace{.7cm} 
	\rho = 
	\frac{1}{2} (p_y - p_z) \;,
\end{equation}
or
\begin{equation}\label{nullrhominus}
	\rho = 
	-\frac{1}{32\pi} \Bigg[ \left( \frac{\partial\beta}{\partial y} \right)^2 
	+ \left( \frac{\partial\beta}{\partial z} \right)^2\Bigg] \hspace{.7cm} \longrightarrow \hspace{.7cm} 
	\rho = 
	\frac{1}{2} (p_z - p_y) \;, 
\end{equation}

where we have used condition (\ref{cons-3}) together with equations (\ref{al-6}) and (\ref{cons-5}). Due to the restrictions imposed by Einstein's equations the energy density for the Alcubierre warp metric is unequivocally dependent of the tangential anisotropies. 

As can be observed in equations (\ref{nullrhoplus}) and (\ref{nullrhominus}), 
the sign of $\rho$ will depend on which of the tangential pressures dominates.

Another result that follows from the constraints found is that the normal pressure $p_x$ must also depend on the anisotropy in general, that is
\begin{equation}
	-\frac{3}{4}  \left( \frac{\partial\beta}{\partial z} \right)^2 
	= 8\pi p_x 
	\hspace{.7cm} \longrightarrow \hspace{.7cm}  
	\frac{3}{2} (p_y - p_z) = p_x = 3\rho \;. \label{pxani1} 
\end{equation}
or
\begin{equation}
	-\frac{3}{4}  \left( \frac{\partial\beta}{\partial y} \right)^2 
	= 8\pi p_x 
	\hspace{.7cm} \longrightarrow \hspace{.7cm}  
	\frac{3}{2} (p_z - p_y) = p_x = 3\rho \;. \label{pxani2} 
\end{equation}
There are equations of state imposed by the field equations that completely determine the behavior of the system and leave room to remove the exotics signs appearing in the matter content.

It is interesting to examine what happens when we consider another class of fluids. For example, when we have a partially anisotropic fluid with $p_x=p$ and $p_y=p_z = p_\perp$, we obtain from equations (\ref{pxani1}) and (\ref{pxani2})
\begin{eqnarray}
	p = 0\;, \hspace{.5cm} \rho = 0 
	\hspace{.7cm} \longrightarrow \hspace{.7cm}  
	\frac{\partial\beta}{\partial z}
	= 0
	\;, \label{pxani1-1} \\
	p = 0\;, \hspace{.5cm} \rho = 0 
	\hspace{.7cm} \longrightarrow \hspace{.7cm}  
	\frac{\partial\beta}{\partial y}
	= 0
	\;, \label{pxani2-1}
\end{eqnarray}
that is, both the normal pressure $p$ and the density $\rho$ cancel out. The same phenomenon occurs if we consider an isotropic fluid or dust. Thus we find that the occurrence of a non-zero energy density necessarily depends on the system being totally anisotropic. The same is valid if an isotropic fluid where $p_x = p_y = p_z = p$ or even dust with $p_x = p_y = p_z = 0$ is considered. In either case, equations (\ref{pxani1-1}) and (\ref{pxani2-1}) are reproduced.
This fact is a consequence of taking into account all the Einstein equations and their implications on the material content. We stress the fact that this result is completely general.

\section{Warp drive in cylindrical coordinates along the $z$--direction}\label{sec-03}

Following an analogous procedure to \cite{Abellan2023,Bobrick:2021wog}, we analyse the spacetime geometry and its consequence using cylindrical coordinates. 

Using the line element (\ref{al-1}) as a guide we constructed the following line element in cylindrical coordinates 
\begin{eqnarray}\label{line-ele}
ds^2 &=& -dt^2 + dr^2 + r^2d\varphi^2 + (dz - \beta dt)^2 \nonumber \\
&=& -(1-\beta^2)dt^2 - 2\beta dt dz
+ dr^2 + r^2d\varphi^2 + dz^2 \,. \;\;\;\;\;
\end{eqnarray}
Here, we have warped in the $z$-direction. To write this expression we have transformed the residual flat space that appears in the Alcubierre warp metric into cylindrical coordinates. As in the previous section, for us the form function $\beta=\beta(t,r,\varphi,z)$ is arbitrary and we will rather look for those conditions that restrict its form from the field equations.

From the line element (\ref{line-ele}) we can write the metric explicitly for the $z$-warp in cylindrical coordinates
\begin{equation}\label{metric}
g_{\mu\nu} =
\left[ \begin{array}{cccc}
	-(1-\beta^2) & 0 & 0 & -\beta \\
	0 & 1 & 0 & 0 \\
	0 & 0 & r^2 & 0 \\
	-\beta & 0 & 0 & 1
\end{array} \right] .
\end{equation}
We can notice that this metric has an analogous form to the Alcubierre metric. With the warp bubble moving in the $z$ direction and a flat residual metric expressed in cylindrical coordinates. This metric admits a Killing vector $\xi_\mu = (-1,0,0,0)$ if $\beta$ is time-independent and therefore the spacetime is stationary.

We are interested in providing matter content in the form of a energy--momentum tensor so that we can study whether the spacetime proposed by Alcubierre (described here in cylindrical coordinates) can be sustained using this matter configuration.

In order to write the energy--momentum tensor, we consider an Eulerian observer described by the following 4-velocity
\begin{equation}\label{alc-4}
u_\mu = (-1,0,0,0)\;, \hspace{.7cm}
u^\mu = (1,0,0,\beta)\;.
\end{equation}
Where $u_\mu $ is a timelike vector obeying $u^\mu u_\mu = -1$. Using this parameterisation, we consider a completely anisotropic fluid given by the expression
\begin{equation}\label{temunuani}
T_{\mu \nu}=(\rho + p_\varphi)u_\mu u_\nu + p_\varphi g_{\mu\nu} + (p_z - p_\varphi)s_\mu s_\nu + (p_r - p_\varphi) t_\mu t_\nu \;,
\end{equation}
where $u_\mu$ is given by (\ref{alc-4}) and $s_\mu$ and $t_\mu$ are
\begin{equation}\label{4vectors}
s_\mu = (-\beta,0,0,1) \;, \hspace{.7cm}
t_\mu = (0,1,0,0)\;.
\end{equation}
These vectors are spacelike and satisfy the relations $s^\mu s_\mu = 1$, $t^\mu t_\mu = 1$, $s^\mu t_\mu = 0$, $u^\mu s_\mu = 0$, and $u^\mu t_\mu = 0$. In matrix form we have
\begin{equation}\label{temunu-matrix-cil}
T_{\mu\nu} =
\left[ \begin{array}{cccc}
	\rho + \beta^2 p_z & 0 & 0 & -\beta p_z \\
	0 & p_r & 0 & 0 \\
	0 & 0 & r^2 p_\varphi & 0 \\
	-\beta p_z & 0 & 0 & p_z
\end{array} \right] .
\end{equation}
In these expressions $\rho$ corresponds to the energy density, $p_z$ is the normal pressure in $z$ direction, and $p_r$ and $p_\varphi$ are the tangential pressures in $r$ and $\varphi$ directions respectively. If we want to constrain to isotropic matter we put $p_r = p_\varphi = p_z = p$, and if we want to consider dust we use $p_i = 0$ for all $i$. We can also discuss a partially anisotropic case which will be relevant below and where $p_r = p_\varphi$ has to be examined.

Assuming a form function $\beta=\beta(t,r,\varphi,z)$ and using the metric (\ref{metric}) and the energy--momentum tensor (\ref{temunu-matrix-cil}) we can write Einstein's equations (\ref{e-eq}) for this geometry. After simplifying we find
\begin{eqnarray}
-\frac{1}{4r^2} \left[ \left( \frac{\partial\beta}{\partial \varphi} \right)^2 +
r^2 \left( \frac{\partial\beta}{\partial r} \right)^2
\right]
&=& 8\pi \rho \;, \;\;\;\;\;\;\;\;\;\; \label{ei-1} \\
\frac{\partial^2\beta}{\partial r \partial z}
&=& 0 \;,  \label{ei-2} \\
\left( \frac{\partial\beta}{\partial z} \right)^2 
+ \beta \frac{\partial^2\beta}{\partial z^2} 
+ \frac{\partial^2\beta}{\partial t \partial z}
&=& -4\pi (p_r + p_\varphi) \;, \;\;\;\; \label{ei-3} \\
\frac{\partial^2\beta}{\partial \varphi \partial z}
&=& 0 \;,  \label{ei-4} \\
\frac{\partial\beta}{\partial r} \frac{\partial\beta}{\partial \varphi}
&=& 0 \;,  \label{ei-5} \\
-\frac{1}{2r^2} \left[ \left( \frac{\partial\beta}{\partial \varphi} \right)^2 -
r^2 \left( \frac{\partial\beta}{\partial r} \right)^2
\right]
&=& 8\pi (p_r - p_\varphi) \;,   \label{ei-6} \\
\frac{\partial^2\beta}{\partial \varphi^2}
+ r 
\frac{\partial\beta}{\partial r}
+ r^2 \frac{\partial^2\beta}{\partial r^2}
&=& 0 \;,  \label{ei-7} \\
2 \frac{\partial\beta}{\partial r} \frac{\partial\beta}{\partial z}
+ \beta \frac{\partial^2\beta}{\partial r \partial z} 
+ \frac{\partial^2\beta}{\partial t \partial r} 
&=& 0 \;,  \label{ei-8} \\
2 \frac{\partial\beta}{\partial \varphi} 
\frac{\partial\beta}{\partial z}
+ \beta \frac{\partial^2\beta}{\partial \varphi \partial z} 
+ \frac{\partial^2\beta}{\partial t \partial \varphi} 
&=& 0 \;,  \label{ei-9} \\
-\frac{3}{4r^2} \left[ \left( \frac{\partial\beta}{\partial \varphi} \right)^2 +
r^2 \left( \frac{\partial\beta}{\partial r} \right)^2
\right]
&=& 8\pi p_z \;.  \label{ei-10} 
\end{eqnarray}
The equations (\ref{ei-1})--(\ref{ei-10}) constitute a coupled system of non--linear partial differential equations. We do not solve these equations, but rather to make some general remarks that can be derived from them. 
We note the formal similarity in this set of equations with those found after simplifying in the previous section in Cartesian coordinates.
We can see from equation (\ref{ei-5}) that two cases emerge which are worth studying in detail.

\subsection{Case 1: $\displaystyle \frac{\partial\beta}{\partial r} = 0$} \label{sec-3}
When this condition is imposed on the system of equations, we find that it reduces to
\begin{eqnarray}
-\frac{1}{4r^2} \left( \frac{\partial\beta}{\partial \varphi} \right)^2 
&=& 8\pi \rho \;, \;\;\;\;\;\;\;\;\;\;  \label{ei-11} \\
\left( \frac{\partial\beta}{\partial z} \right)^2 
+ \beta \frac{\partial^2\beta}{\partial z^2} 
+ \frac{\partial^2\beta}{\partial t \partial z}
&=& -4\pi (p_r + p_\varphi) \;,  \label{ei-13} \\
\frac{\partial^2\beta}{\partial \varphi \partial z} &=& 0 \;,  \label{ei-14}\\     
-\frac{1}{2r^2} \left( \frac{\partial\beta}{\partial \varphi} \right)^2 
&=& 8\pi (p_r - p_\varphi) \;,   \label{ei-16} \\
\frac{\partial^2\beta}{\partial \varphi^2}
&=& 0 \;,  \label{ei-17} \\
2 \frac{\partial\beta}{\partial \varphi} 
\frac{\partial\beta}{\partial z}
+ \beta \frac{\partial^2\beta}{\partial \varphi \partial z} 
+ \frac{\partial^2\beta}{\partial t \partial \varphi} 
&=& 0 \;,  \label{ei-19} \\
-\frac{3}{4r^2} \left( \frac{\partial\beta}{\partial \varphi} \right)^2
&=& 8\pi p_z \;.  \label{ei-110} 
\end{eqnarray}
We recognise that by requiring the form function to be of type $\beta=\beta(t,\varphi,z)$, the system is significantly reduced. We can also observe from the equations (\ref{ei-11}), (\ref{ei-16}) and (\ref{ei-110}) that some relations naturally arise for the components of the energy--momentum tensor
\begin{eqnarray}
p_z &=& 3\rho \;, \label{eos-11} \\
2\rho &=& p_r - p_\varphi \;. \label{eos-12}
\end{eqnarray}
These relations help to characterise the nature of the fluid that could support the warp metric.
Even further, since equation (\ref{ei-14}) is also a constrain it implies the possibility that $\beta$ is independent either of $z$ or $\varphi$. If $\beta$ does not depend on $z$ we have
\begin{equation}
p_r = -p_\varphi \;. \label{eos-13}
\end{equation}
The case when $\beta$ is independent of $\varphi$ leads to the results in section \ref{sec-4}.

\subsection{Case 2: $\displaystyle \frac{\partial\beta}{\partial \varphi} = 0$} \label{sec-4}
Now, evaluating the second possible condition for the equation (\ref{ei-5}), we obtain
\begin{eqnarray}
-\frac{1}{4} \left( \frac{\partial\beta}{\partial r} \right)^2 
&=& 8\pi \rho \;, \;\;\;\;\;\;\;\;\;\;  \label{ei-21} \\
\left( \frac{\partial\beta}{\partial z} \right)^2 
+ \beta \frac{\partial^2\beta}{\partial z^2} 
+ \frac{\partial^2\beta}{\partial t \partial z}
&=& -4\pi (p_r + p_\varphi) \;,  \label{ei-23} \\
\frac{\partial^2\beta}{\partial r \partial z} &=& 0 \;, 
\label{ei-24}\\   
\frac{1}{2} \left( \frac{\partial\beta}{\partial r} \right)^2 
&=& 8\pi (p_r - p_\varphi) \;,   \label{ei-26} \\
\frac{\partial\beta}{\partial r}
+ r \frac{\partial^2\beta}{\partial r^2}
&=& 0 \;,  \label{ei-27} \\
2 \frac{\partial\beta}{\partial r} \frac{\partial\beta}{\partial z}
+ \beta \frac{\partial^2\beta}{\partial r \partial z} 
+ \frac{\partial^2\beta}{\partial t \partial r} 
&=& 0 \;,  \label{ei-28} \\
-\frac{3}{4} \left( \frac{\partial\beta}{\partial r} \right)^2
&=& 8\pi p_z \;.  \label{ei-210} 
\end{eqnarray}
Once more the constraint permits reducing the system of equations by considering $\beta = \beta(t,r,z)$. In addition, relations emerge between the functions describing the matter content, this time are given by
\begin{eqnarray}
p_z &=& 3\rho \;, \label{eos-21} \\
2 \rho &=& p_\varphi - p_r \; \label{eos-22}
\end{eqnarray}
We note that the equation relating $\rho$ and $p_z$ is the same as in the previous case but the relation between $\rho$ and the tangential pressures changes sign. As in the previous case, in order to fulfil equation (\ref{ei-24}) the function $\beta$ should be either independent of $r$ or $z$. The $r$--independent case leads to the results of the previous section \ref{sec-3}. In addition, when $\beta$ does not depend on $z$ the equation (\ref{eos-13}) is recovered.

So far we have written down the simplified Einstein equations and found that they impose a restriction on the form function $\beta$, namely it cannot depend on both $r$, $\varphi$ and also on $z$.

Looking at equation (\ref{ei-1}) we see that it is essentially the negative energy result reported from Alcubierre's seminal paper. 

It is important to remark that, in reality, the appearance of both terms is forbidden by the constrain (\ref{ei-5}), consequently, either (\ref{ei-11}) or (\ref{ei-21}) must be satisfied.

With this symmetry, and in the same way as in Cartesian coordinates, there is an interesting result when the tangential pressures are equal, $p_\varphi = p_r = p_\perp$. By imposing this condition on cases 1 and 2, we see that they reduce to a single case described by the following equations
\begin{eqnarray}
\rho &=& 0 \;,  \label{ei-31} \\
p_z &=& 0 \;,  \label{ei-32} \\
\left( \frac{\partial\beta}{\partial z} \right)^2 
+ \beta \frac{\partial^2\beta}{\partial z^2} 
+ \frac{\partial^2\beta}{\partial t \partial z}
&=& -8\pi p_\perp \;.  \label{ei-33} 
\end{eqnarray}
In this special case we obtain a fluid that has only tangential equal pressures and acts as a source for spacetime dynamics. In terms of the derivatives of $\beta$, this is equivalent to saying that it is independent of $r$ and $\varphi$ coordinates. The remarkable thing is that the density cancels identically and therefore there is no negative energy problem.

The results of our analysis reveal that in systems such as dust, isotropic fluid, or anisotropic fluids with equal tangential pressures, the problem of negative energy does not arise.
In other words, the negative energy problem is necessarily associated with the complete anisotropy of the system.

\subsection{Energy density analysis}\label{sec-5}
The results obtained above for cases 1 and 2 provide us with a new expression for the energy density equation.
Here we can find an interesting relation when considering either (\ref{ei-11}), (\ref{eos-12}) or (\ref{ei-21}), (\ref{eos-22})
\begin{eqnarray}
\mbox{Case 1:}& \hspace{.7cm} \displaystyle
-\frac{1}{4r^2} \left( \frac{\partial\beta}{\partial \varphi} \right)^2 
= 4\pi (p_r - p_\varphi) \;, \\
\mbox{Case 2:}& \hspace{.7cm} \displaystyle
-\frac{1}{4} \left( \frac{\partial\beta}{\partial r} \right)^2 
= 4\pi (p_\varphi - p_r) \;.
\end{eqnarray}
In either case it can be seen that the right-hand side can be negative depending on which of the tangential pressures dominates. On the other hand, we might consider what happens if the beta function is also independent of the $z$ coordinate
\begin{eqnarray}
\mbox{Case 1:}& \hspace{.3cm} \rho = -p_\varphi\;, \hspace{.7cm}
\displaystyle
\frac{1}{4r^2} \left( \frac{\partial\beta}{\partial \varphi} \right)^2 
= 8\pi p_\varphi \;, \\
\mbox{Case 2:}& \hspace{.3cm} \rho = -p_r\;, \hspace{.7cm}
\displaystyle
\frac{1}{4} \left( \frac{\partial\beta}{\partial r} \right)^2 
= 8\pi  p_r \;.
\end{eqnarray}
Again, it is observed that the expression for the density is closely related to the anisotropy of the system and in both cases no inconsistencies are evident. In particular, it is not observed that the energy density must be negative. Finally, for the partially anisotropic case, if $\beta$ does not depend on the $z$ coordinate, the matter content must be zero, $\rho=0$, $p_z=p_r=p_\varphi=0$.

This case turns out to be very similar to the original Alcubierre case discussed in the previous section with the obvious advantage that its analysis and the drawing of consequences is much simpler.

\section{Warp drive in cylindrical coordinates along the $r$--direction}\label{sec-04}
Another way of approaching the warp drive using cylindrical coordinates is to implement it in the radial $r$-coordinate.
In this case the metric is written as
\begin{equation}\label{metric-r-cyl}
g_{\mu\nu} =
\left[ \begin{array}{cccc}
	-(1-\beta^2) & -\beta & 0 & 0 \\
	-\beta & 1 & 0 & 0 \\
	0 & 0 & r^2 & 0 \\
	0 & 0 & 0 & 1
\end{array} \right] .
\end{equation}
Following a procedure analogous to that of the previous section, we can write the momentum energy tensor for an anisotropic fluid in these coordinates. In this way we find
\begin{equation}\label{temunu-matrix}
T_{\mu\nu} =
\left[ \begin{array}{cccc}
	\rho + \beta^2 p_r & -\beta p_r & 0 & 0 \\
	-\beta p_r & p_r & 0 & 0 \\
	0 & 0 & r^2 p_\varphi & 0 \\
	0 & 0 & 0 & p_z
\end{array} \right] .
\end{equation}
As in the previous sections, this momentum energy tensor can be particularised to consider different cases such as isotropic fluid or dust.

Assuming again a function $\beta(t,r,\varphi,z)$, and after simplifying, we write the Einstein equations for this configuration
\begin{eqnarray}
-\frac{1}{4r^2} \left[ \left( \frac{\partial\beta}{\partial \varphi} \right)^2 +
r^2 \left( \frac{\partial\beta}{\partial z} \right)^2
- 4r\beta \frac{\partial\beta}{\partial r}
\right]
&=& 8\pi \rho \;, \;\;\;\;\;\;\;\;\;\; \label{ei-rcil-1} \\
\frac{\partial^2\beta}{\partial \varphi^2} +
r^2 \frac{\partial^2\beta}{\partial z^2}
&=& 0 \;,  \label{ei-rcil-2} \\
-\frac{1}{4r^2} \left[ 3\!\left( \frac{\partial\beta}{\partial \varphi} \right)^2 +
3r^2 \left( \frac{\partial\beta}{\partial z} \right)^2
+ 4r \left( \beta \frac{\partial\beta}{\partial r}
+ \frac{\partial\beta}{\partial t} \right)
\right]
&=& 8\pi p_r \;, \;\;\;\; \label{ei-rcil-3} \\
\frac{\partial\beta}{\partial\varphi} - r\frac{\partial^2\beta}{\partial r \partial \varphi}
&=& 0 \;,  \label{ei-rcil-4} \\
2r \frac{\partial\beta}{\partial r} 
\frac{\partial\beta}{\partial \varphi}
+ \frac{\partial^2\beta}{\partial t \partial \varphi}
&=& 0 \;,  \label{ei-rcil-5} \\
\frac{1}{4} \left\{
\left(\frac{\partial\beta}{\partial\varphi}\right)^2 -
r^2 \left[ \left( \frac{\partial\beta}{\partial z} \right)^2 +
4\left( \frac{\partial\beta}{\partial r} \right)^2 +
4\beta \frac{\partial^2\beta}{\partial r^2} +
4 \frac{\partial^2\beta}{\partial t \partial r}
\right]
\right\}
&=& 8\pi r^2 p_\varphi \;,   \label{ei-rcil-6} \\
\frac{\partial\beta}{\partial z} + r\frac{\partial^2\beta}{\partial r \partial z}
&=& 0 \;,  \label{ei-rcil-7} \\
2 \frac{\partial\beta}{\partial r} \frac{\partial\beta}{\partial z}
+ \frac{\partial^2\beta}{\partial t \partial z} 
&=& 0 \;,  \label{ei-rcil-8} \\
\frac{\partial\beta}{\partial \varphi}
\frac{\partial\beta}{\partial z}
&=& 0 \;,  \label{ei-rcil-9} \\
\frac{1}{2} \left[ 
r^2 \left( \frac{\partial\beta}{\partial z} \right)^2 -
\left( \frac{\partial\beta}{\partial \varphi} \right)^2 -
2r \left( 2\beta \frac{\partial\beta}{\partial r} +
\frac{\partial\beta}{\partial t}
\right)
\right]
&=& 8\pi r^2 (p_z - p_\varphi) \;.  \label{ei-rcil-10} 
\end{eqnarray}
It is important to note that this set of equations has a very different structure from the case studied in the previous section where we examined a cylindrical warp in the $z$-direction. In fact in the present case it is not trivial to find a relationship purely in terms of the anisotropic fluid variables.

\subsection{Case 1: $\displaystyle \frac{\partial\beta}{\partial z}=0$}
It is clear from the above equations that by using the expression (\ref{ei-rcil-9}) we can simplify the whole set and write a reduced and more neat set of equations. In this sense, considering $\beta$ independent of the $z$-coordinate we obtain
\begin{eqnarray}
-\frac{1}{4r^2} \left[ \left( \frac{\partial\beta}{\partial \varphi} \right)^2
- 4r\beta \frac{\partial\beta}{\partial r}
\right]
&=& 8\pi \rho \;, \;\;\;\;\;\;\;\;\;\; \label{ei-rcil-21} \\
\frac{\partial^2\beta}{\partial \varphi^2}
&=& 0 \;,  \label{ei-rcil-22} \\
-\frac{1}{4r^2} \left[ 3\!\left( \frac{\partial\beta}{\partial \varphi} \right)^2
+ 4r \left( \beta \frac{\partial\beta}{\partial r}
+ \frac{\partial\beta}{\partial t} \right)
\right]
&=& 8\pi p_r \;, \;\;\;\; \label{ei-rcil-23} \\
\frac{\partial\beta}{\partial\varphi} - r\frac{\partial^2\beta}{\partial r \partial \varphi}
&=& 0 \;,  \label{ei-rcil-24} \\
2r \frac{\partial\beta}{\partial r} 
\frac{\partial\beta}{\partial \varphi}
+ \frac{\partial^2\beta}{\partial t \partial \varphi}
&=& 0 \;,  \label{ei-rcil-25} \\
\frac{1}{4} \left\{
\left(\frac{\partial\beta}{\partial\varphi}\right)^2 -
4r^2 \left[
\left( \frac{\partial\beta}{\partial r} \right)^2 +
\beta \frac{\partial^2\beta}{\partial r^2} +
\frac{\partial^2\beta}{\partial t \partial r}
\right]
\right\}
&=& 8\pi r^2 p_\varphi \;,   \label{ei-rcil-26} \\
-\frac{1}{2} \left[ 
\left( \frac{\partial\beta}{\partial \varphi} \right)^2 +
2r \left( 2\beta \frac{\partial\beta}{\partial r} +
\frac{\partial\beta}{\partial t}
\right)
\right]
&=& 8\pi r^2 (p_z - p_\varphi) \;.  \label{ei-rcil-27} 
\end{eqnarray}
Moreover, we can find a relation between the fluid variables as
\begin{equation}\label{eos-rcil-01}
\rho = p_r + p_\varphi - p_z \;.
\end{equation}
In addition, if we consider that the function $\beta$ is also independent of $r$, we observe by equation (\ref{ei-rcil-24}) that it must necessarily also be independent of $\varphi$. As a consequence, the density $\rho$ and the tangential pressure $p_\varphi$ vanish, leaving only the expression
\begin{equation}\label{eos-rcil-02}
p_z = p_r \;.
\end{equation}
This corresponds to an anisotropic fluid model with a peculiar anisotropy configuration.

\subsection{Case 2: $\displaystyle \frac{\partial\beta}{\partial \varphi}=0$}
In this case, using the fact that the function $\beta$ is independent of the $\varphi$-coordinate, the equations are simplified to
\begin{eqnarray}
-\frac{1}{4r^2} \left[ r^2 \left( \frac{\partial\beta}{\partial z} \right)^2
- 4r\beta \frac{\partial\beta}{\partial r}
\right]
&=& 8\pi \rho \;, \;\;\;\;\;\;\;\;\;\; \label{ei-rcil-31} \\
\frac{\partial^2\beta}{\partial z^2}
&=& 0 \;,  \label{ei-rcil-32} \\
-\frac{1}{4r^2} \left[ 3r^2\!\left( \frac{\partial\beta}{\partial z} \right)^2
+ 4r \left( \beta \frac{\partial\beta}{\partial r}
+ \frac{\partial\beta}{\partial t} \right)
\right]
&=& 8\pi p_r \;, \;\;\;\; \label{ei-rcil-33} \\
-\frac{1}{4}
\left[ \left(\frac{\partial\beta}{\partial z}\right)^2 +
4\left( \frac{\partial\beta}{\partial r} \right)^2 +
4\beta \frac{\partial^2\beta}{\partial r^2} +
4\frac{\partial^2\beta}{\partial t \partial r}
\right]
&=& 8\pi p_\varphi \;,   \label{ei-rcil-34} \\
\frac{\partial\beta}{\partial z} + \frac{\partial^2\beta}{\partial r \partial z}
&=& 0 \;,  \label{ei-rcil-35} \\
2 \frac{\partial\beta}{\partial r} 
\frac{\partial\beta}{\partial z}
+ \frac{\partial^2\beta}{\partial t \partial z}
&=& 0 \;,  \label{ei-rcil-36} \\
\frac{1}{2} \left[ 
r^2 \left( \frac{\partial\beta}{\partial z} \right)^2 -
2r \left( 2\beta \frac{\partial\beta}{\partial r} +
\frac{\partial\beta}{\partial t}
\right)
\right]
&=& 8\pi r^2 (p_z - p_\varphi) \;.  \label{ei-rcil-37} 
\end{eqnarray}
A simple relationship between the fluid variables cannot be obtained in this scenario. The same will occur in the next section where we deal with spherical symmetry. 
Now, considering that the function $\beta$ does not depend on the $r$-coordinate, we see by (\ref{ei-rcil-35}) that it is necessarily also independent of the $z$-coordinate. Thus we observe that $\rho=0$, $p_\varphi=0$ and the relation (\ref{eos-rcil-02}) of the previous case is again satisfied.

What is important to mention is that both expressions for density (\ref{ei-rcil-21}) and (\ref{ei-rcil-31}), do not necessarily imply that density $\rho$ must be negative as is often claimed. Moreover, we find again that non--zero densities depend on the system being completely anisotropic as we have seen in the cases studied previously.

\section{Warp Drive in Spherical Coordinates}\label{sec-05}

In a prior article, we were able to formulate a warp metric with spherical coordinates using the Alcubierre metric as a basis \cite{Abellan2023}.
We propose the following metric
\begin{equation}\label{metric-sph}
g_{\mu\nu} =
\left[ \begin{array}{cccc}
	-(1-\beta^2) & -\beta & 0 & 0 \\
	-\beta & 1 & 0 & 0 \\
	0 & 0 & r^2 & 0 \\
	0 & 0 & 0 & r^2\sin^2{\!\theta}
\end{array} \right] .
\end{equation}
This corresponds to a type of Alcubierre's metric with a warp bubble on radial direction and the residual flat space described by spherical coordinates. In the following we will explore the conditions on $\beta$ due to the constrains imposed by the field equations.

In general, we consider the form function as a quantity dependent of both time and radial coordinates $\beta(t,r)$.

\subsection{Anisotropic matter content in the spherical warp}\label{polyaniso}

In order to study the matter content, we consider an Eulerian observer as it was done in the previous sections discribing an anisotropic fluid in the form

\begin{equation}
T_{\mu\nu}=\left[ \begin{array}{cccc}
	\rho + \beta^2 p_r & -\beta p_r & 0 & 0\\
	-\beta p_r & p_r & 0& 0\\
	0 & 0 & r^2\, p_\perp &0 \\
	0 &0 & 0& r^2 \sin^2\!\theta\, p_\perp
\end{array} \right] .
\end{equation}

This corresponds to an anisotropic fluid, which reduces to isotropic case when $p_\perp = p_r$. An interesting aspect with respect to the previous cases is that due to the spherical symmetry, both tangential pressures are necessarily equal. This can be seen by examining the Einstein equations related to $G_{\theta\theta}$ and $G_{\varphi\varphi}$. 

Using the metric and the energy--momentum tensor we write the Einstein equations for this system
\begin{eqnarray}
\frac{\beta}{r^2} \left( \beta + 2r \de{\beta}{r} \right) \!\!&=&\!\! 8\pi \rho\;, \label{cosmic-01} \\
\frac{\beta}{r^2} \left(\beta + 2r \frac{\partial \beta}{\partial r}\right) + \frac{2}{r} \frac{\partial\beta}{\partial t}  \!\!&=&\!\! -8\pi p_r \;, \label{cosmic-02} \\
\beta^2 + r \frac{\partial\beta}{\partial t} - r^2 \left[\de{}{r}\left(\beta \de{\beta}{r}\right) + \frac{\partial^2 \beta}{\partial t \partial r} \right] \!\!&=&\!\! 8\pi r^2 \Delta \;, \label{cosmic-03}  \;\;\;\;\;
\end{eqnarray}
here $\Delta = p_\perp - p_r$ is the anisotropy factor. This system differs significantly from the cases discussed above. In the present case there is no simple relation between the physical quantities of the fluid. What can be clearly seen is that there is no condition that requires the energy to be negative. The system of equations obtained has as sources the density $\rho$, the radial pressure $p_r$ and the anisotropy factor $\Delta$.

In summary, it can be stated that: (i) the energy density is in general finite, even in the absence of anisotropy, differing with the Alcubierre's original warp. (ii) The sign of $\rho$ is not restricted to be negative, it depends by construction on positive and negative terms and furthermore the anisotropy by definition can have negative or positive contributions that add to overall the density.

\section{Conclusions}\label{sec-06}
In this work we have carried out an exhaustive revision of the Alcubierre--type metrics with general $\beta$ function, considering the Einstein equations in their entirety. We have considered some modifications to the geometry of the original metric and evaluated some of the consequences on the matter content exploring warps using cylindrical and spherical coordinates. By performing a detailed inspection of the equations we have managed to express them in a convenient way so that it is possible to extract simple constraints on the beta function and thus draw conclusions for the material content, in particular for the energy density.

The main conclusion of this work is that the assertion that the Alcubierre warp requires by necessity negative energy density should be taken with caution. In particular, we have seen how the matter content, expressed in the momentum energy tensor, modifies the possible values of the energy density.

In the specific case of the Alcubierre warp in Cartesian coordinates we have seen how the density depends directly on the difference in tangential pressures and therefore of the complete anisotropy of the fluid. Thus a fluid with equal tangential pressures, a perfect isotropic fluid and even dust produce null energy densities. Moreover, in the case where the tangential pressures are different, the energy density cannot necessarily be said to be negative. The dependence of the energy density on the difference in tangential pressures allows us to have a mechanism by which this quantity would effectively reach negative values and not as a characterisation of an exotic fluid type.

When exploring the warp metric in other symmetries we have found qualitatively different behaviour in each of the metrics considered. The geometrical configuration most similar to the original Cartesian case is the one corresponding to a warp in cylindrical coordinates in the $z$-direction. We obtain a completely analogous relation between the anisotropic fluid variables in this case, and observe the dependence of the density on the difference of the tangential pressures.

When considering the cylindrical warp in the $r$-direction and the spherical warp in the $r$-direction, it is observed that the results change. In these cases it is often not even possible to find a relationship between the physical variables of the fluid itself. However, the most important thing is that in both cases the expression of the energy density does not prove that it must be negative as has been recurrently claimed in the literature.

The claims of negative energy density have been made from studying the expression relating the energy density to the geometry (\ref{al-6}), omitting all the restrictions that are imposed by the rest of the Einstein equations. Moreover, it is only recently that we have begun to study what the appropriate material content should be in order to support the geometry of warp drives and thus understand their real physical feasibility.

In this sense, we believe that further studies should be undertaken to help elucidate the appropriate matter content that gives rise to warp configurations. In this way we will gain an understanding not only of the geometrical aspects of the theory but also of the nature of the matter that makes them viable.

\section*{Declarations}

%
%
\begin{itemize}
\item Funding\\
Not applicable
\item Conflict of interest/Competing interests \\
The authors declare that they have no competing interests as defined by Springer, or other interests that might be perceived to influence the results and/or discussion reported in this paper.

\item Ethics approval \\
Not applicable
\item Consent to participate\\
Not applicable
\item Consent for publication\\
Not applicable
\item Availability of data and materials\\
The manuscript does not contain any material from third parties; all of the material is
owned by the authors and/or no permissions are required.
\item Code availability \\
Not applicable
\item Authors' contributions\\
The authors equally contributed to the conceptualization and analysis. Writing of the
manuscript by NB and GA.
\end{itemize}
%

%
%
%



\bibliography{bibio}


\begin{thebibliography}{34}
\ifx \bisbn   \undefined \def \bisbn  #1{ISBN #1}\fi
\ifx \binits  \undefined \def \binits#1{#1}\fi
\ifx \bauthor  \undefined \def \bauthor#1{#1}\fi
\ifx \batitle  \undefined \def \batitle#1{#1}\fi
\ifx \bjtitle  \undefined \def \bjtitle#1{#1}\fi
\ifx \bvolume  \undefined \def \bvolume#1{\textbf{#1}}\fi
\ifx \byear  \undefined \def \byear#1{#1}\fi
\ifx \bissue  \undefined \def \bissue#1{#1}\fi
\ifx \bfpage  \undefined \def \bfpage#1{#1}\fi
\ifx \blpage  \undefined \def \blpage #1{#1}\fi
\ifx \burl  \undefined \def \burl#1{\textsf{#1}}\fi
\ifx \doiurl  \undefined \def \doiurl#1{\url{https://doi.org/#1}}\fi
\ifx \betal  \undefined \def \betal{\textit{et al.}}\fi
\ifx \binstitute  \undefined \def \binstitute#1{#1}\fi
\ifx \binstitutionaled  \undefined \def \binstitutionaled#1{#1}\fi
\ifx \bctitle  \undefined \def \bctitle#1{#1}\fi
\ifx \beditor  \undefined \def \beditor#1{#1}\fi
\ifx \bpublisher  \undefined \def \bpublisher#1{#1}\fi
\ifx \bbtitle  \undefined \def \bbtitle#1{#1}\fi
\ifx \bedition  \undefined \def \bedition#1{#1}\fi
\ifx \bseriesno  \undefined \def \bseriesno#1{#1}\fi
\ifx \blocation  \undefined \def \blocation#1{#1}\fi
\ifx \bsertitle  \undefined \def \bsertitle#1{#1}\fi
\ifx \bsnm \undefined \def \bsnm#1{#1}\fi
\ifx \bsuffix \undefined \def \bsuffix#1{#1}\fi
\ifx \bparticle \undefined \def \bparticle#1{#1}\fi
\ifx \barticle \undefined \def \barticle#1{#1}\fi
\bibcommenthead
\ifx \bconfdate \undefined \def \bconfdate #1{#1}\fi
\ifx \botherref \undefined \def \botherref #1{#1}\fi
\ifx \url \undefined \def \url#1{\textsf{#1}}\fi
\ifx \bchapter \undefined \def \bchapter#1{#1}\fi
\ifx \bbook \undefined \def \bbook#1{#1}\fi
\ifx \bcomment \undefined \def \bcomment#1{#1}\fi
\ifx \oauthor \undefined \def \oauthor#1{#1}\fi
\ifx \citeauthoryear \undefined \def \citeauthoryear#1{#1}\fi
\ifx \endbibitem  \undefined \def \endbibitem {}\fi
\ifx \bconflocation  \undefined \def \bconflocation#1{#1}\fi
\ifx \arxivurl  \undefined \def \arxivurl#1{\textsf{#1}}\fi
\csname PreBibitemsHook\endcsname

\bibitem{Alcubierre:1994tu}
\begin{barticle}
\bauthor{\bsnm{Alcubierre}, \binits{M.}}:
\batitle{{The Warp drive: Hyperfast travel within general relativity}}.
\bjtitle{Class. Quant. Grav.}
\bvolume{11},
\bfpage{73}--\blpage{77}
(\byear{1994})
{\href{https://arxiv.org/abs/gr-qc/0009013}{{arXiv:gr-qc/0009013}}}.
\doiurl{10.1088/0264-9381/11/5/001}
\end{barticle}
\endbibitem

\bibitem{Visser1995lwet}
\begin{bbook}
\bauthor{\bsnm{{Visser}}, \binits{M.}}:
\bbtitle{Lorentzian Wormholes. From Einstein to Hawking},
(\byear{1995})
\end{bbook}
\endbibitem

\bibitem{Everett:1995nn}
\begin{barticle}
\bauthor{\bsnm{Everett}, \binits{A.E.}}:
\batitle{{Warp drive and causality}}.
\bjtitle{Phys. Rev. D}
\bvolume{53},
\bfpage{7365}--\blpage{7368}
(\byear{1996}).
\doiurl{10.1103/PhysRevD.53.7365}
\end{barticle}
\endbibitem

\bibitem{Krasnikov:1995ad}
\begin{barticle}
\bauthor{\bsnm{Krasnikov}, \binits{S.V.}}:
\batitle{{Hyperfast travel in general relativity}}.
\bjtitle{Phys. Rev. D}
\bvolume{57},
\bfpage{4760}--\blpage{4766}
(\byear{1998})
{\href{https://arxiv.org/abs/gr-qc/9511068}{{arXiv:gr-qc/9511068}}}.
\doiurl{10.1103/PhysRevD.57.4760}
\end{barticle}
\endbibitem

\bibitem{Hiscock:1997ya}
\begin{barticle}
\bauthor{\bsnm{Hiscock}, \binits{W.A.}}:
\batitle{{Quantum effects in the Alcubierre warp drive space-time}}.
\bjtitle{Class. Quant. Grav.}
\bvolume{14},
\bfpage{183}--\blpage{188}
(\byear{1997})
{\href{https://arxiv.org/abs/gr-qc/9707024}{{arXiv:gr-qc/9707024}}}.
\doiurl{10.1088/0264-9381/14/11/002}
\end{barticle}
\endbibitem

\bibitem{Everett:1997hb}
\begin{barticle}
\bauthor{\bsnm{Everett}, \binits{A.E.}},
\bauthor{\bsnm{Roman}, \binits{T.A.}}:
\batitle{{A Superluminal subway: The Krasnikov tube}}.
\bjtitle{Phys. Rev. D}
\bvolume{56},
\bfpage{2100}--\blpage{2108}
(\byear{1997})
{\href{https://arxiv.org/abs/gr-qc/9702049}{{arXiv:gr-qc/9702049}}}.
\doiurl{10.1103/PhysRevD.56.2100}
\end{barticle}
\endbibitem

\bibitem{Visser:1997tq}
\begin{barticle}
\bauthor{\bsnm{Visser}, \binits{M.}}:
\batitle{General relativistic energy conditions: The hubble expansion in the
  epoch of galaxy formation}.
\bjtitle{Phys. Rev. D}
\bvolume{56},
\bfpage{7578}--\blpage{7587}
(\byear{1997})
{\href{https://arxiv.org/abs/gr-qc/9705070}{{arXiv:gr-qc/9705070}}}.
\doiurl{10.1103/PhysRevD.56.7578}
\end{barticle}
\endbibitem

\bibitem{Olum:1998mu}
\begin{barticle}
\bauthor{\bsnm{Olum}, \binits{K.D.}}:
\batitle{{Superluminal travel requires negative energies}}.
\bjtitle{Phys. Rev. Lett.}
\bvolume{81},
\bfpage{3567}--\blpage{3570}
(\byear{1998})
{\href{https://arxiv.org/abs/gr-qc/9805003}{{arXiv:gr-qc/9805003}}}.
\doiurl{10.1103/PhysRevLett.81.3567}
\end{barticle}
\endbibitem

\bibitem{Low:1998uy}
\begin{barticle}
\bauthor{\bsnm{Low}, \binits{R.J.}}:
\batitle{{Speed limits in general relativity}}.
\bjtitle{Class. Quant. Grav.}
\bvolume{16},
\bfpage{543}--\blpage{549}
(\byear{1999})
{\href{https://arxiv.org/abs/gr-qc/9812067}{{arXiv:gr-qc/9812067}}}.
\doiurl{10.1088/0264-9381/16/2/016}
\end{barticle}
\endbibitem

\bibitem{Pfenning:1998ua}
\begin{botherref}
\oauthor{\bsnm{Pfenning}, \binits{M.J.}}:
{Quantum inequality restrictions on negative energy densities in curved
  space-times}.
Other thesis
(April 1998)
\end{botherref}
\endbibitem

\bibitem{Gonzalez-Diaz:1999vtr}
\begin{barticle}
\bauthor{\bsnm{Gonzalez-Diaz}, \binits{P.F.}}:
\batitle{{Warp drive space-time}}.
\bjtitle{Phys. Rev. D}
\bvolume{62},
\bfpage{044005}
(\byear{2000})
{\href{https://arxiv.org/abs/gr-qc/9907026}{{arXiv:gr-qc/9907026}}}.
\doiurl{10.1103/PhysRevD.62.044005}
\end{barticle}
\endbibitem

\bibitem{Visser:1999de}
\begin{bchapter}
\bauthor{\bsnm{Visser}, \binits{M.}},
\bauthor{\bsnm{Barcelo}, \binits{C.}}:
\bctitle{{Energy conditions and their cosmological implications}}.
In: \bbtitle{3rd International Conference on Particle Physics and the Early
  Universe},
pp. \bfpage{98}--\blpage{112}
(\byear{2000}).
\doiurl{10.1142/9789812792129_0014}
\end{bchapter}
\endbibitem

\bibitem{Clark:1999zn}
\begin{barticle}
\bauthor{\bsnm{Clark}, \binits{C.}},
\bauthor{\bsnm{Hiscock}, \binits{W.A.}},
\bauthor{\bsnm{Larson}, \binits{S.L.}}:
\batitle{{Null geodesics in the Alcubierre warp drive space-time: The View from
  the bridge}}.
\bjtitle{Class. Quant. Grav.}
\bvolume{16},
\bfpage{3965}--\blpage{3972}
(\byear{1999})
{\href{https://arxiv.org/abs/gr-qc/9907019}{{arXiv:gr-qc/9907019}}}.
\doiurl{10.1088/0264-9381/16/12/313}
\end{barticle}
\endbibitem

\bibitem{VanDenBroeck:1999sn}
\begin{barticle}
\bauthor{\bsnm{Van Den~Broeck}, \binits{C.}}:
\batitle{{A 'Warp drive' with reasonable total energy requirements}}.
\bjtitle{Class. Quant. Grav.}
\bvolume{16},
\bfpage{3973}--\blpage{3979}
(\byear{1999})
{\href{https://arxiv.org/abs/gr-qc/9905084}{{arXiv:gr-qc/9905084}}}.
\doiurl{10.1088/0264-9381/16/12/314}
\end{barticle}
\endbibitem

\bibitem{VanDenBroeck:1999xs}
\begin{botherref}
\oauthor{\bsnm{Van Den~Broeck}, \binits{C.}}:
{On the (im)possibility of warp bubbles}
(1999)
{\href{https://arxiv.org/abs/gr-qc/9906050}{{arXiv:gr-qc/9906050}}}
\end{botherref}
\endbibitem

\bibitem{Barcelo:2000zf}
\begin{barticle}
\bauthor{\bsnm{Barcelo}, \binits{C.}},
\bauthor{\bsnm{Visser}, \binits{M.}}:
\batitle{{Scalar fields, energy conditions, and traversable wormholes}}.
\bjtitle{Class. Quant. Grav.}
\bvolume{17},
\bfpage{3843}--\blpage{3864}
(\byear{2000})
{\href{https://arxiv.org/abs/gr-qc/0003025}{{arXiv:gr-qc/0003025}}}.
\doiurl{10.1088/0264-9381/17/18/318}
\end{barticle}
\endbibitem

\bibitem{Barcelo:2002bv}
\begin{barticle}
\bauthor{\bsnm{Barcelo}, \binits{C.}},
\bauthor{\bsnm{Visser}, \binits{M.}}:
\batitle{{Twilight for the energy conditions?}}
\bjtitle{Int. J. Mod. Phys. D}
\bvolume{11},
\bfpage{1553}--\blpage{1560}
(\byear{2002})
{\href{https://arxiv.org/abs/gr-qc/0205066}{{arXiv:gr-qc/0205066}}}.
\doiurl{10.1142/S0218271802002888}
\end{barticle}
\endbibitem

\bibitem{Lobo:2002zf}
\begin{barticle}
\bauthor{\bsnm{Lobo}, \binits{F.}},
\bauthor{\bsnm{Crawford}, \binits{P.}}:
\batitle{{Weak energy condition violation and superluminal travel}}.
\bjtitle{Lect. Notes Phys.}
\bvolume{617},
\bfpage{277}--\blpage{291}
(\byear{2003})
{\href{https://arxiv.org/abs/gr-qc/0204038}{{arXiv:gr-qc/0204038}}}
\end{barticle}
\endbibitem

\bibitem{Lobo:2004an}
\begin{bchapter}
\bauthor{\bsnm{Lobo}, \binits{F.S.N.}},
\bauthor{\bsnm{Visser}, \binits{M.}}:
\bctitle{{Linearized warp drive and the energy conditions}}.
In: \bbtitle{27th Spanish Relativity Meeting: Beyond General Relativity (ERE
  2004)}
(\byear{2004})
\end{bchapter}
\endbibitem

\bibitem{Lobo:2004wq}
\begin{barticle}
\bauthor{\bsnm{Lobo}, \binits{F.S.N.}},
\bauthor{\bsnm{Visser}, \binits{M.}}:
\batitle{{Fundamental limitations on 'warp drive' spacetimes}}.
\bjtitle{Class. Quant. Grav.}
\bvolume{21},
\bfpage{5871}--\blpage{5892}
(\byear{2004})
{\href{https://arxiv.org/abs/gr-qc/0406083}{{arXiv:gr-qc/0406083}}}.
\doiurl{10.1088/0264-9381/21/24/011}
\end{barticle}
\endbibitem

\bibitem{Curiel:2014zba}
\begin{barticle}
\bauthor{\bsnm{Curiel}, \binits{E.}}:
\batitle{{A Primer on Energy Conditions}}.
\bjtitle{Einstein Stud.}
\bvolume{13},
\bfpage{43}--\blpage{104}
(\byear{2017})
{\href{https://arxiv.org/abs/1405.0403}{{arXiv:1405.0403}}}
{[physics.hist-ph]}.
\doiurl{10.1007/978-1-4939-3210-8_3}
\end{barticle}
\endbibitem

\bibitem{Lobo2017}
\begin{botherref}
\oauthor{\bsnm{{Lobo}}, \binits{F.S.N.}}:
{Wormholes, Warp Drives and Energy Conditions}.
Fundamental Theories of Physics
\textbf{189}
(2017).
\doiurl{10.1007/978-3-319-55182-1}
\end{botherref}
\endbibitem

\bibitem{Alcubierre:2017kqf}
\begin{barticle}
\bauthor{\bsnm{Alcubierre}, \binits{M.}},
\bauthor{\bsnm{Lobo}, \binits{F.S.N.}}:
\batitle{{Warp Drive Basics}}.
\bjtitle{Fundam. Theor. Phys.}
\bvolume{189},
\bfpage{257}--\blpage{279}
(\byear{2017}).
\doiurl{10.1007/978-3-319-55182-1_11}
\end{barticle}
\endbibitem

\bibitem{Kontou:2020bta}
\begin{barticle}
\bauthor{\bsnm{Kontou}, \binits{E.-A.}},
\bauthor{\bsnm{Sanders}, \binits{K.}}:
\batitle{Energy conditions in general relativity and quantum field theory}.
\bjtitle{Class. Quant. Grav.}
\bvolume{37}(\bissue{19}),
\bfpage{193001}
(\byear{2020})
{\href{https://arxiv.org/abs/2003.01815}{{arXiv:2003.01815}}}
{[gr-qc]}.
\doiurl{10.1088/1361-6382/ab8fcf}
\end{barticle}
\endbibitem

\bibitem{Bobrick:2021wog}
\begin{barticle}
\bauthor{\bsnm{Bobrick}, \binits{A.}},
\bauthor{\bsnm{Martire}, \binits{G.}}:
\batitle{{Introducing Physical Warp Drives}}.
\bjtitle{Class. Quant. Grav.}
\bvolume{38}(\bissue{10}),
\bfpage{105009}
(\byear{2021})
{\href{https://arxiv.org/abs/2102.06824}{{arXiv:2102.06824}}}
{[gr-qc]}.
\doiurl{10.1088/1361-6382/abdf6e}
\end{barticle}
\endbibitem

\bibitem{Lentz:2021rzh}
\begin{bchapter}
\bauthor{\bsnm{Lentz}, \binits{E.W.}}:
\bctitle{{Hyper-Fast Positive Energy Warp Drives}}.
In: \bbtitle{{16th Marcel Grossmann Meeting on~Recent Developments in
  Theoretical and Experimental General Relativity, Astrophysics and
  Relativistic Field Theories}}
(\byear{2021})
\end{bchapter}
\endbibitem

\bibitem{Santiago:2021aup}
\begin{barticle}
\bauthor{\bsnm{Santiago}, \binits{J.}},
\bauthor{\bsnm{Schuster}, \binits{S.}},
\bauthor{\bsnm{Visser}, \binits{M.}}:
\batitle{{Generic warp drives violate the null energy condition}}.
\bjtitle{Phys. Rev. D}
\bvolume{105}(\bissue{6}),
\bfpage{064038}
(\byear{2022})
{\href{https://arxiv.org/abs/2105.03079}{{arXiv:2105.03079}}}
{[gr-qc]}.
\doiurl{10.1103/PhysRevD.105.064038}
\end{barticle}
\endbibitem

\bibitem{Schuster:2022ati}
\begin{botherref}
\oauthor{\bsnm{Schuster}, \binits{S.}},
\oauthor{\bsnm{Santiago}, \binits{J.}},
\oauthor{\bsnm{Visser}, \binits{M.}}:
{ADM mass in warp drive spacetimes}
(2022)
{\href{https://arxiv.org/abs/2205.15950}{{arXiv:2205.15950}}}
{[gr-qc]}
\end{botherref}
\endbibitem

\bibitem{Santos-Pereira:2020puq}
\begin{barticle}
\bauthor{\bsnm{Santos-Pereira}, \binits{O.L.}},
\bauthor{\bsnm{Abreu}, \binits{E.M.C.}},
\bauthor{\bsnm{Ribeiro}, \binits{M.B.}}:
\batitle{Dust content solutions for the alcubierre warp drive spacetime}.
\bjtitle{Eur. Phys. J. C}
\bvolume{80}(\bissue{8}),
\bfpage{786}
(\byear{2020})
{\href{https://arxiv.org/abs/2008.06560}{{arXiv:2008.06560}}}
{[gr-qc]}.
\doiurl{10.1140/epjc/s10052-020-8355-2}
\end{barticle}
\endbibitem

\bibitem{Santos-Pereira:2021mqp}
\begin{barticle}
\bauthor{\bsnm{Santos-Pereira}, \binits{O.L.}},
\bauthor{\bsnm{Abreu}, \binits{E.M.C.}},
\bauthor{\bsnm{Ribeiro}, \binits{M.B.}}:
\batitle{Perfect fluid warp drive solutions with the cosmological constant}.
\bjtitle{Eur. Phys. J. Plus}
\bvolume{136}(\bissue{9}),
\bfpage{902}
(\byear{2021})
{\href{https://arxiv.org/abs/2108.10960}{{arXiv:2108.10960}}}
{[gr-qc]}.
\doiurl{10.1140/epjp/s13360-021-01899-7}
\end{barticle}
\endbibitem

\bibitem{Santos-Pereira:2021mrr}
\begin{barticle}
\bauthor{\bsnm{Santos-Pereira}, \binits{O.L.}},
\bauthor{\bsnm{Abreu}, \binits{E.M.C.}},
\bauthor{\bsnm{Ribeiro}, \binits{M.B.}}:
\batitle{Charged dust solutions for the warp drive spacetime}.
\bjtitle{Gen. Rel. Grav.}
\bvolume{53}(\bissue{2}),
\bfpage{23}
(\byear{2021})
{\href{https://arxiv.org/abs/2102.05119}{{arXiv:2102.05119}}}
{[gr-qc]}.
\doiurl{10.1007/s10714-021-02799-y}
\end{barticle}
\endbibitem

\bibitem{Santos-Pereira:2021rsr}
\begin{bchapter}
\bauthor{\bsnm{Santos-Pereira}, \binits{O.L.}},
\bauthor{\bsnm{Abreu}, \binits{E.M.C.}},
\bauthor{\bsnm{Ribeiro}, \binits{M.B.}}:
\bctitle{{Warp drive dynamic solutions considering different fluid sources}}.
In: \bbtitle{{16th Marcel Grossmann Meeting on~Recent Developments in
  Theoretical and Experimental General Relativity, Astrophysics and
  Relativistic Field Theories}}
(\byear{2021})
\end{bchapter}
\endbibitem

\bibitem{Santos-Pereira:2021sdm}
\begin{barticle}
\bauthor{\bsnm{Santos-Pereira}, \binits{O.L.}},
\bauthor{\bsnm{Abreu}, \binits{E.M.C.}},
\bauthor{\bsnm{Ribeiro}, \binits{M.B.}}:
\batitle{Fluid dynamics in the warp drive spacetime geometry}.
\bjtitle{Eur. Phys. J. C}
\bvolume{81}(\bissue{2}),
\bfpage{133}
(\byear{2021})
{\href{https://arxiv.org/abs/2101.11467}{{arXiv:2101.11467}}}
{[gr-qc]}.
\doiurl{10.1140/epjc/s10052-021-08921-3}
\end{barticle}
\endbibitem

\bibitem{Abellan2023}
\begin{barticle}
\bauthor{\bsnm{Abell{\'a}n}, \binits{G.}},
\bauthor{\bsnm{Bolivar}, \binits{N.}},
\bauthor{\bsnm{Vasilev}, \binits{I.}}:
\batitle{Alcubierre warp drive in spherical coordinates with some matter
  configurations}.
\bjtitle{The European Physical Journal C}
\bvolume{83}(\bissue{1}),
\bfpage{7}
(\byear{2023}).
\doiurl{10.1140/epjc/s10052-022-11091-5}
\end{barticle}
\endbibitem

\end{thebibliography}
\end{document}